\documentclass[12pt,a4paper]{cibb}
\usepackage{subfigure,graphicx}
\usepackage{amsmath,amsfonts,latexsym,amssymb,euscript,xr}
\usepackage{cite}

\title{\large $\ $\\ \bf USING THE LASSO FOR GENE SELECTION IN BLADDER CANCER DATA}

\author{ St\'ephane Chr\'etien$^{(1)}$, Christophe Guyeux$^{(2)}$, 
Michael Boyer-Guittaut,
R\'egis Delage-Mouroux$^{(3)}$
and Fran\c coise Desc\^otes$^{(4)}$}
\bigskip
\address{$(1)$ Laboratoire de Math\'ematiques de Besan\c con, Universit\'e de Franche-Comt\'e,
	16, route de Gray, 25000 Besan\c{c}on, France. \\
$(2)$~FEMTO-ST Institute, UMR 6174 CNRS, DISC Computer Science Department 
	Universit\'e de Franche-Comt\'e,
	16, route de Gray, 25000 Besan\c{c}on, France. \\
	$(3)$ EA 3922/IFR133
Universit\'e de Franche-Comt\'e - UFR Sciences et Techniques EA 3922/IFR133 - 25030 Besan\c con. \\
	$(4)$ Service de Biochimie et Biologie Mol\'eculaire Sud, Pavillon 3D, Centre Hospitalier Lyon Sud, Pierre B\'enite Cedex 69495, France.\\
\emph{Corresponding author:stephane.chretien@univ-fcomte.fr}}

\abstract{bladder cancer, gene marker, generalized LASSO, gaussian mixtures, model selection, clustering.
\\[17pt]
{\bf Abstract.} Given a gene expression data array of a list of bladder cancer patients with their tumor states, it may be difficult to determine which genes can operate as disease markers when the array is large and possibly contains outliers and missing data. An additional difficulty is that  observations (tumor states) in 
the regression problem are discrete ones. In this article, we solve these problems on concrete data using first a clustering approach, followed by Least Absolute Shrinkage and Selection Operator (LASSO) estimators in a nonlinear regression problem
involving discrete variables,
as described in the brand-new research work of Plan and Vershynin. Gene markers of the most severe tumor state are finally provided using the proposed approach.}

\begin{document}
\thispagestyle{myheadings}
\pagestyle{myheadings}
\markright{\tt Proceedings of CIBB 2015}


\section{\bf Introduction}

In this article, we present a methodology to perform selection among genes based 
on their expression in various groups of patients, in order to find new genetic markers for 
specific pathologies. Our approach is based on clustering the denoised data and computing a LASSO (Least Absolute Shrinkage and Selection Operator) estimator, 
in order to select the relevant genes. This latter 
belongs to the class of penalized regression estimators where the penalty is  a multiple of the 
$\ell_1$-norm of the regression vector. 

We apply the proposed methodology to a set of gene expression data from patients with bladder cancer, where
four possible subtypes of tumor state are considered thus making the observations discrete in 
the regression problem. Our primary objective in the present work is to extract a set of relevant 
genes for the bladder cancer under study. 
A secondary objective is to emphasize the following fact: although the regression problem we consider is nonlinear 
and involves discrete variables, the LASSO can still be used if selection is performed for prediction. This is due to the recent work of Plan and Vershynin~\cite{Plan}, which merits 
advertising in applications to biology where much of methodology concentrates on the binary logistic 
model and too often neglects more complicated outputs\footnote{For instance, the data in our study can be 
modelled using ordered polytomic regression and needs a penalized likelihood estimator. This latter is hard to find in current statistical software libraries whereas the LASSO is widely available.}. 

The remainder of this article is as follows. The bladder cancer data that have been used for illustration purpose in this study are described in the next section.
Recalls about Gaussian mixture and selection model are 
provided in Section~\ref{sec:clustering} together with a
principal component analysis (PCA) of the data under study.
The generalized LASSO of Vershynin and Plan is recalled and applied to our data in Section~\ref{sec:lasso}.
This research work ends by a conclusion that proposes a gene marker for the last tumor state.

\section{\bf Presentation of the data}

To accurately diagnose bladder cancer is a Public Health priority as for instance, in 2013, 10,000 new patients were affected by this cancer in France. A promising way to improve
the diagnosis and to design efficient treatments is to determine which genes are responsible for such a cancer. 
Obviously, such medical treatments must depend on 
how much the tumor is developed.
To achieve this goal, gene expression data together with 
corresponding state of the malignant tumor in the bladder
have been recently collected from 100 patients in the Lyon 
region, France~\cite{ISBRA15}.

\begin{figure}[h]
\vspace{3mm}
\centering
  \subfigure[2 components]{\includegraphics[width=5cm]{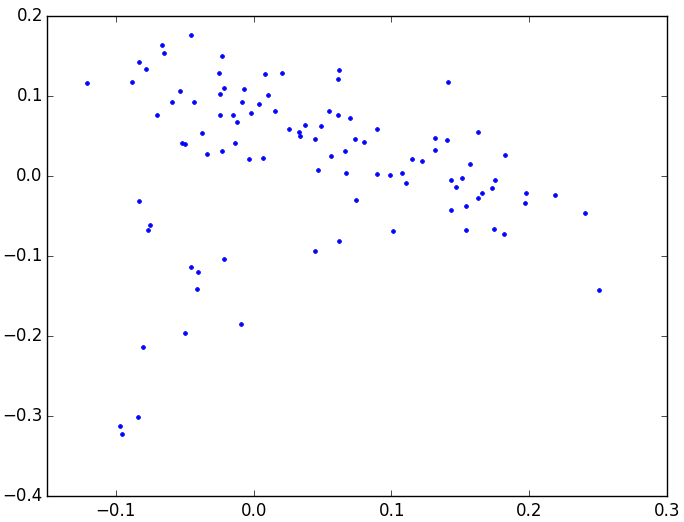}}
   \subfigure[3 components]{\includegraphics[width=6cm]{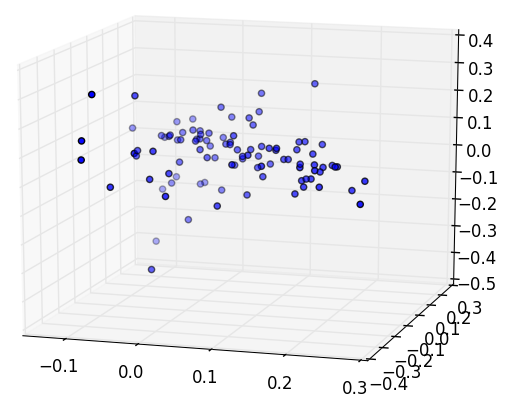}} 
\caption {Principal component analyses of the denoised data\label{fig:apcNB}}
\vspace{-8mm}
\end{figure}

In this study, 
34 genes of interest have been chosen, while the tumor state has been decomposed in 4 classes, namely: 
\begin{itemize}
\item $Ta$ : noninfiltrating tumor in Urothelium; 
\item $T1a$ : noninfiltrating tumor in Urothelium and parts of the chorion;
\item $T1b$ : noninfiltrating tumor in Urothelium and the full chorion;
\item $>T1$ : infiltrating tumor. 
\end{itemize}
Remark that, in the standard classification, the last group of the list above incorporates states from $T2$ to $T4b$. 
Mathematically speaking, the data are thus constituted by this first column providing the tumor state (discrete), 
followed by 34 other columns that quantify the expression of the selected genes. Each row of the $100\times 35$ matrix is associated to one of the 100 given patients, and the objective is to determine, for each tumor state, which gene(s) must be selected as the best marker candidate(s). However, due to its experimental origin, this raw array contains outliers and corrupted data.

To extract the relevant features in a given dataset is a difficult task, recently resolved in the non-negative data case with the 
Non-negative Matrix factorization (NMF) method.
The objective of our previous research work~\cite{ISBRA15} was to extend this method to the case of missing and/or corrupted data due to outliers.
To do so, data have been denoised, missing values have been imputed, and outliers
have been detected while performing a low-rank non-negative matrix factorization of
the recovered matrix. To achieve this goal, a mixture of Bregman proximal methods and of
the Augmented Lagrangian scheme have been used on our dataset in~\cite{ISBRA15}, in a similar way to the so-called Alternating Direction of Multipliers method. In what follows, we thus deal with two arrays: the
raw one and the denoised one. A principal component analysis of this latter is provided in Figure~\ref{fig:apcNB}. Next stages consist of determining if these denoised data can be clusterized well (which is not the case for the raw data), if the optimal number of clusters corresponds well to the number of tumor states, and if each cluster is coherent, that is, if each patient being in tumor state $k$ is too in cluster number $k$.

\section{\bf Clustering of the gene expression data}
\label{sec:clustering}
\subsection{Gaussian mixtures}\label{gmm}
Finite Gaussian mixture models (GMM) are widely used in a great number of application fields as a mean to perform model based classification. From pattern recognition to biology, from quality control to finance, many examples have shown the pertinence of the Gaussian mixture model approach~\cite{FMM}. 
In GMM data $Y_1,\ldots,Y_n$ are assumed independent and identically distributed (i.i.d.) and to be drawn from  the density:
\begin{equation}
\sum_{k=1}^K p^*_k f^{(d)}(y;\mu_k,\Sigma_k)
\end{equation}
where 
\begin{eqnarray}
f^{(d)}(y;\mu,\Sigma) & = & 
\frac1{\sqrt{(2\pi)^d {\rm det} (\Sigma^*)\big)}} \exp \Big( -\frac12 (y-\mu^*)^t{\Sigma^*}^{-1}(y-\mu^*) \Big)
\end{eqnarray}
and where the vector $\theta^*=(p_1^*,\ldots,p_K^*,\mu_1^*,\ldots,\mu_K^*,\Sigma_1^*,\ldots,\Sigma_k^*)$ is an unknown multidimensional parameter. To this model, 
we traditionally associate an extended model using the notion of complete data. In mixture models, the complete data are i.i.d. couples of the form $(Y_i,Z_i)$, where $Z_i$ is a multinomial random variable taking values in $\{1,\ldots,K\}$ with $P(Z_i=k)=p^*_k$.
This latter represents the index of the mixture component from which observation $i$ was drawn. We assume that, conditionally 
on the event $Z_i=k$, $Y_i$ has density 
\begin{align}
\frac1{\sqrt{(2\pi)^d {\rm det} (\Sigma_k^*)}} \exp \Big( -\frac12 (y-\mu_k^*)^t{\Sigma_k^*}^{-1}(y-\mu_k^*) \Big).
\end{align} 
The variables $Z_1,\ldots,Z_n$ being unobserved, they are usually called ``latent variables''. 
The standard approach for estimating $\theta^*$ is the maximum likelihood methodology that consists of finding $\hat{\theta}$ which maximizes the log-likelihood function 
\begin{equation}
\label{lkhd}
l(\theta)=\sum_{i=1}^n \log \Big( \sum_{k=1}^K p_k f^{(d)}(y;\mu_k,\Sigma_k) \Big)   
\end{equation}
over the set 
\begin{equation*}
\Theta = \Big\{(p_1,...,p_K,\mu_1,...,\mu_K,\Sigma_1,...,\Sigma_K) \mid p_k \in \mathbb R_+, \: 
\mu_k\in \mathbb R^d,\: \Sigma_k \in \mathbb S_d^+, \textrm{ and }\sum_{k=1}^K p_k=1 \Big\}
\end{equation*} 
where $\mathbb S_d^+$ denotes the set 
of all symmetric positive semidefinite matrices and $\mathbb R_+$ is the set of nonnegative real numbers.
The usual way to maximize the log-likelihood is the so-called EM algorithm~\cite{DempsterEtAl,FMM}, or 
its more efficient componentwise variants, \emph{e.g.}, \cite{FesslerHero,CeleuxChretien}.

\subsection{Results}

The problem of choosing the number of clusters $K$ \emph{a priori} is a difficult one. This is usually done by 
comparing the penalized maximum likelihood values for different values of $K$ and choosing the maximum one. Model selection can be performed using the Bayesian Information Criterion (BIC). This criterion 
is the opposite of the maximum likelihood value penalized with $\log(n)\times$ the number of real parameters to estimate. 

\begin{figure}[h]
\vspace{3mm}
 \begin{center}
 \subfigure[Using all components]{\includegraphics[width=5.5cm]{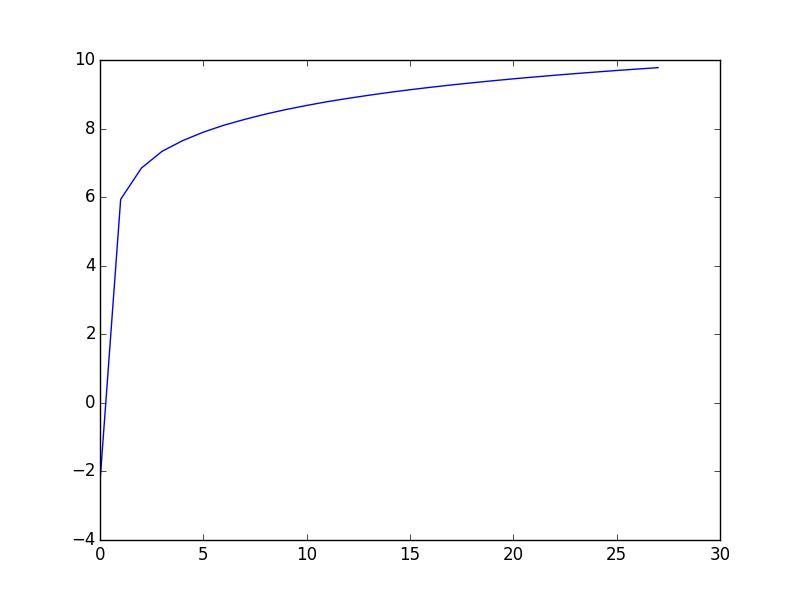}}
 \subfigure[On the 3 principal components]{\includegraphics[width=5.5cm]{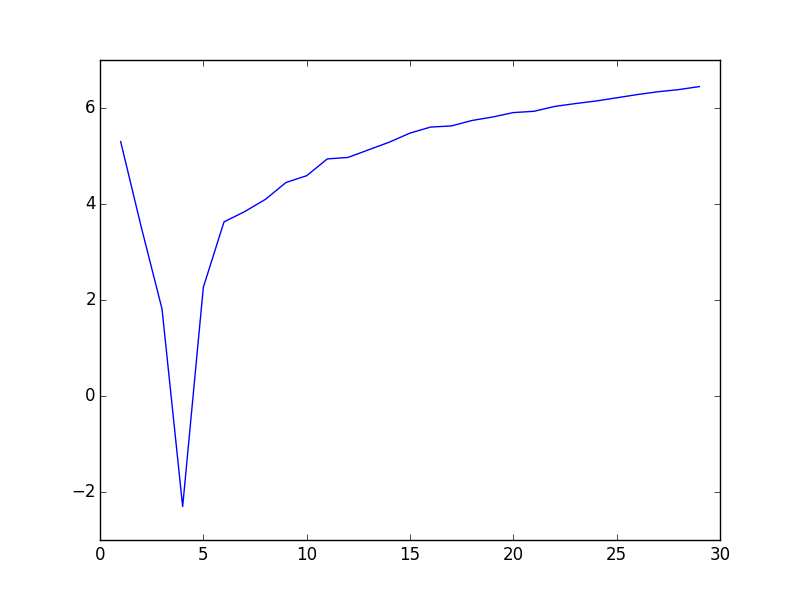}}
\caption {Determination of the optimal number of clusters in denoized data: number of clusters for easting values and (log of) Bayesian Information Criterion (BIC) for northing ones.\label{fig:bic}}
 \end{center}
\vspace{-8mm}
\end{figure}

\begin{figure}[h]
\vspace{3mm}
\centering
 \subfigure[First view]{\includegraphics[width=7cm]{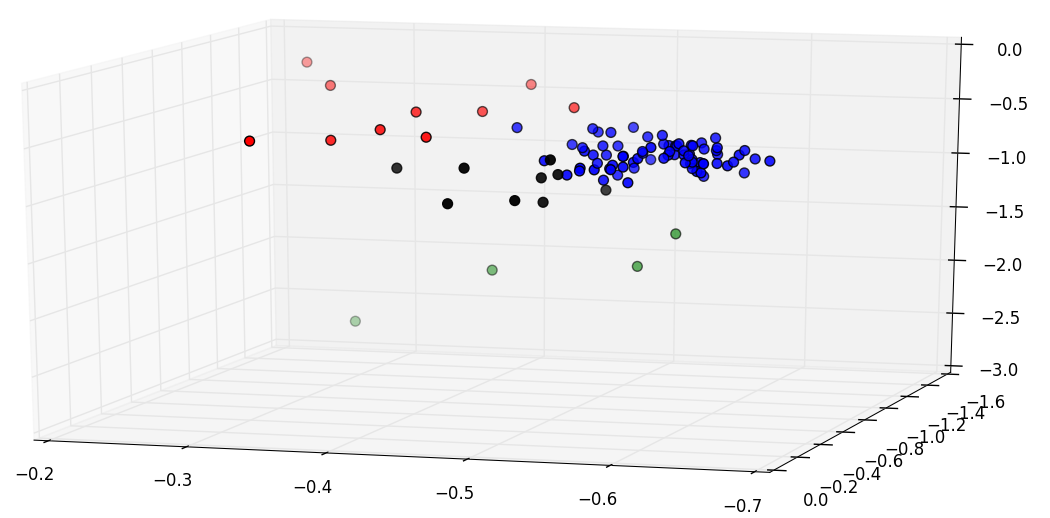}} 
  \subfigure[Second view]{\includegraphics[width=7cm]{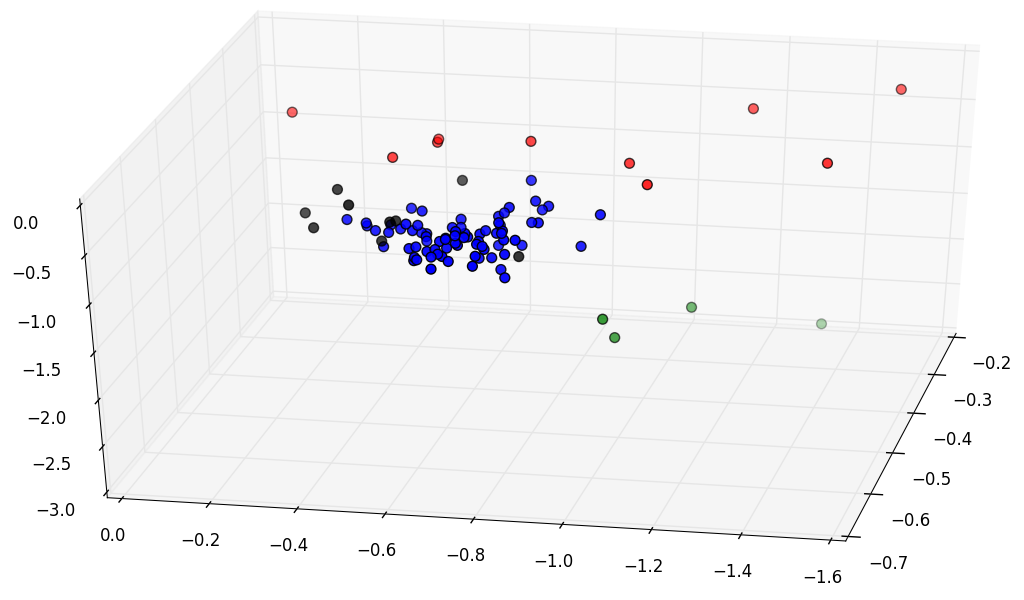}} 
\caption {PCA on raw data, colorized according to their cluster provided by the GMM.\label{fig:apc}}
\vspace{-8mm}
\end{figure}

The first attempt on raw data failed to provide any useful information, due to outliers and missing data. This criterion has then been applied on the gene expression part of our denoised array, to determine the best way to cluster the set of genes. The number of mixture components has ranged from 1 to 29, and at each time the Bayesian information criterion for the current model fit has been computed (more precisely, for pretty prints, the logarithm of $x-minBIC$, where $minBIC$ is the smallest obtained BIC). As can be seen in obtained plot depicted in Figure~\ref{fig:bic}, the criterion has not provide any obvious result when considering the whole data. However, applying it on the 3 principal components shown in Figure~\ref{fig:apcNB} emphasizes that the optimal number of clusters is 4. Such a result were encouraging, as we have 4 tumor states in the array. We then have performed a PCA on the raw data while colorizing each of the 4 clusters provided by the Gaussian mixture model. Obtained results are depicted in Figure~\ref{fig:apc}, they are coherent with the tumor state of each patient. Note that, at this stage, each cluster is a Gaussian one.

\section{\bf Generalized LASSO of Vershynin and Plan}
\label{sec:lasso}
\subsection{Background}

In our data set, we have to explain the tumor state based on the expression of 
certain genes. Since the state is a discrete variable, an appropriate model should be the 
multinomial logistic model. This model could be refined and a more appropriate one could 
be the ordered polytomic regression model. 
The main difficulty in this kind of model 
is that when the number of observations is of the same order as the number of covariates, 
the maximum likelihood estimator might not perform well. Moreover, selection of some of the 
most relevant covariates might be an important way to extract meaningful information from the data. 
Performing variable selection can be done using the Bayesian Information Criterion that 
we presented in Section~\ref{gmm}. However, if one wants to try all models built 
on the expression of less than, \emph{e.g.}, 15 genes, the computational effort might be overwhelming in 
many gene expression studies. 

In order to overcome this issue, an important contribution was made by Tibshirani~\cite{CandesPlanAnnStat09}. 
The analysis 
was then extended to the case of even more covariates than 
observations in~\cite{CandesPlanAnnStat09}, while extension to Generalized Linear Models was then proposed, see~\cite{BuhlmannVanDeGeer} for a very useful 
and mathematically deep reference. 
The LASSO can be described as the solution of the following penalized least squares problem
\begin{align}
\label{optimizationObj}
\min_{b\in \mathbb R^p} \quad \frac12 \Vert y-Xb\Vert_2^2+\lambda \Vert b \Vert_1,
\end{align}
where $\Vert b\Vert_1$ is the $\ell_1$-norm of $b$, \emph{i.e.}, $\Vert b\Vert_1=\sum_{j=1=^p \vert b_j\vert}$ and: $y$ is the vector containing the $n$ outputs (the tumor state), 
 $X$ is the matrix whose columns are the expression of each of the $p$ genes in the $n$ patients,
while $\lambda$ is the relaxation parameter.

Under certain properties of the matrix $X$, the LASSO estimator enjoys good variable recovery properties 
\cite[Thm 1.4]{CandesPlanAnnStat09}. If the matrix has very correlated columns, variable recovery 
will generally fail but under a mixture model, good prediction bounds can still be obtained
as proven in \cite{ChretienMixLasso}. Since in our study, we suspect that high correlations exist between the genes 
under study, we cannot expect to obtain good variable recovery with the LASSO. However, we can still 
believe that the variable selection performed with the LASSO is relevant with respect to prediction. 
Various theoretical values for the relaxation parameter $\lambda$ have been proposed in the literature,
see, \emph{e.g.}, \cite{Chatterjee} or \cite{Chichignou}.

\subsection{The generalized LASSO}
An important point to address in our study is the discrete nature of the output vector $y$. However, as stated previously, extending the analysis of the LASSO to the ordered polytomic model seems 
quite difficult and cumbersome to obtain, while useful and robust software is currently not available. One very interesting question is whether the LASSO can still be applied in the nonlinear context where the output (tumor state) is ordered and discrete as in our problem? Fortunately, the answer is yes, as it was recently proven by Plan and Vershynin~\cite{Plan}. Moreover, their analysis applies to more general penalization terms than just the $\ell_1$ norm. One important assumption in~\cite{Plan} is that the design, \emph{i.e.}, $X$, is Gaussian. Therefore, an important step before using this approach is that the data set must be clustered into Gaussian clusters, as it has been achieved via the mixture model in Section~\ref{gmm}. Then the LASSO should be performed clusterwise. In what follows, we only focus on ``black'' cluster ($>T1$ tumor).

The choice of the relaxation parameter $\lambda$ is still an important and difficult problem in practice. In the sequel, we 
restrict our attention to the results obtained using the LARS 
method, which consists of computing the estimator for a continuous set of values of $\lambda$, and we analyze the obtained trajectory in order to select the most important genes.

\section{\bf Conclusion on experimental results}

\begin{figure}[h]
\vspace{3mm}
 \begin{center}
 \includegraphics[width=11cm]{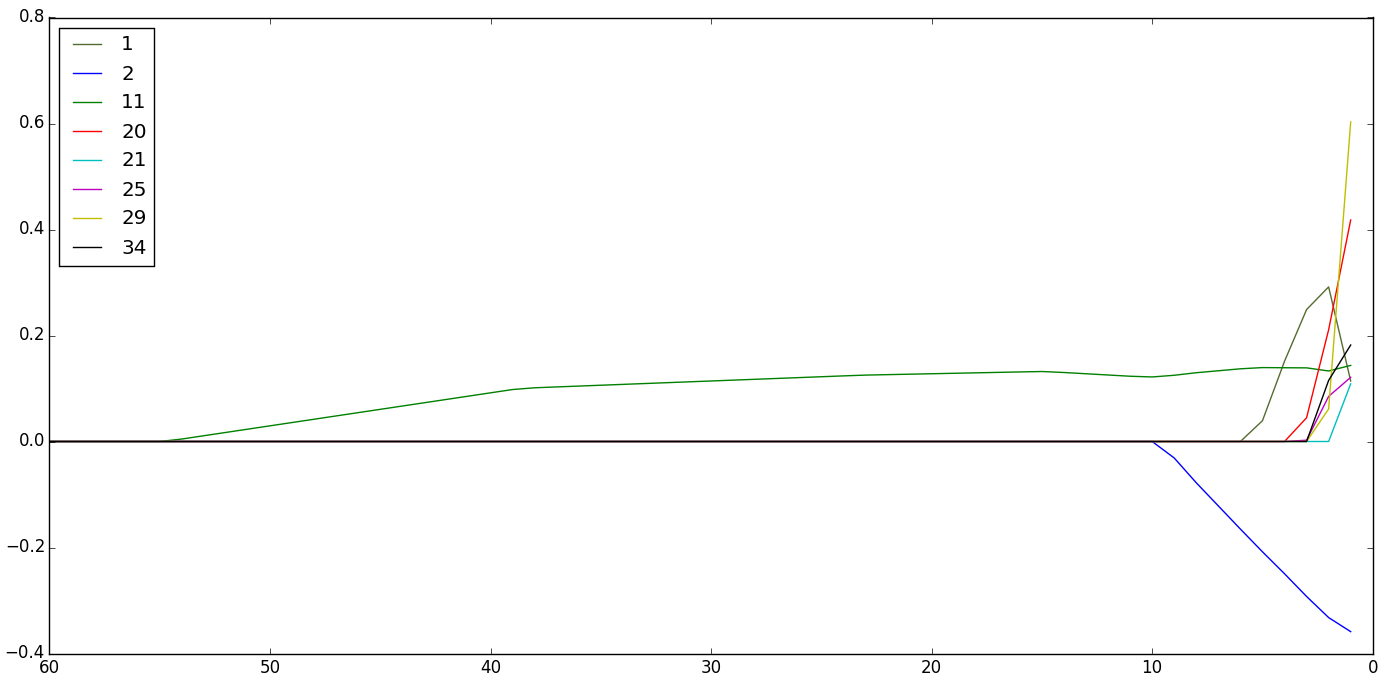}
\caption {Obtained trajectory in the Lars method on infiltrating tumor Gaussian cluster\label{fig:lasso}}
 \end{center}
\vspace{-8mm}
\end{figure}

Eight genes still remain in the Gaussian infiltrating tumor cluster under 
consideration, corresponding in our array to columns: 1 and
2 (genes \emph{ATF3} and \emph{Bcl2 like14} respectively), 11 (\emph{HMGB2}), 20 (\emph{MMP11}), 21 (\emph{ORC6L}), 25 (\emph{RAD54L}), 29 (\emph{TK}), and 34 (\emph{Vimentin}). $\lambda$ value of eq.~\eqref{optimizationObj} ranged between $0.1$ and $0.03$ with a step of $0.001$, and
for each $\lambda$ the penalized least square problem has
been solved. Each resolution has led to the associated parameter vector 
$b \in \mathbb{R}^8$, where the $k$-th coordinate of $b$ 
corresponds to the $k$-th gene of the considered cluster,
according to its column occurrence in the raw
data array. We thus have plotted the LASSO result trajectories, putting
$\lambda$ in the abscissa axis and each coordinate of $b$ in
the ordinate axis, see 
Figure~\ref{fig:lasso}.

As can be seen, HMGB2 (column number 11) obviously steps out of line, appearing as the most important gene reflecting the infiltrating tumor cluster. This is not surprising, as the
high mobility group box 2 (HMGB2) overexpression has been observed in several human tumor types, and is involved in cancer progression and prognosis, especially in the bladder one~\cite{hmgb2,hmgb2bis}. 
Generally speaking, all genes have a positive impact, except the second most important gene, namely \emph{Bcl2 like14} (2), which has an effect diametrically opposed to the other ones. This  can be explained by the well known fact that
overexpression of this gene, a candidate tumor suppressor~\cite{bcl}, induces apoptosis in cells. Finally, by order of importance, the two
next genes are respectively \emph{ATF3} (1) and \emph{MMP11} (20), even though the small value of $\lambda$ may blur
the information raised by these trajectories. 
HMGB2 and these latter can thus act as genetic markers of $>T1$ tumors.



\emph{This work was partially funded by  El\'ements Transposables  Franche-Comt\'e grant.}

\bibliographystyle{apalike}
{\fontsize{10}{10}\selectfont

}
\end{document}